# Quantification of the helical morphology of chiral Au nanorods

Wouter Heyvaert[1], Adrián Pedrazo-Tardajos[1], Ajinkya Kadu[1], Nathalie Claes[1], Guillermo González-Rubio[2,3], Luis M. Liz-Marzán[2,4,5], Wiebke Albrecht[1,6]*, Sara Bals[1]*

[1] EMAT and NANOlab Center of Excellence, University of Antwerp, 2020 Antwerp, Belgium

[2] CIC biomaGUNE, Basque Research and Technology Alliance (BRTA), 20014 Donostia-San Sebastián, Spain

[3] Physical Chemistry Department, University of Konstanz, Universitätsstraße 10, Box 714, 78457 Konstanz, Germany.

[4] Centro de Investigación Biomédica en Red de Bioingeniería Biomateriales, y Nanomedicina (CIBER-BBN), 20014 Donostia-San Sebastián, Spain

[5] Ikerbasque, Basque Foundation for Science, 48009 Bilbao, Spain

[6] Center for Nanophotonics, AMOLF, Science Park 104, 1098 XG Amsterdam, The Netherlands

*Electron Tomography, Chirality, Helicity, Au nanoparticles*

**ABSTRACT:** Chirality in inorganic nanoparticles and nanostructures has gained increasing scientific interest, because of the possibility to tune their ability to interact differently with left- and right-handed circularly polarized light. In some cases, the optical activity is hypothesized to originate from a chiral morphology of the nanomaterial. However, quantifying the degree of chirality in objects with sizes of tens of nanometers is far from straightforward. Electron tomography offers the possibility to faithfully retrieve the three-dimensional morphology of nanomaterials, but only a qualitative interpretation of the morphology of chiral nanoparticles has been possible so far. We introduce herein a methodology that enables us to quantify the helicity of complex chiral nanomaterials, based on the geometrical properties of a helix. We demonstrate that an analysis at the single particle level can provide significant insights into the origin of chiroptical properties.

Chiral features in metal nanoparticles (NP) result in chiroptical properties, of interest to many applications, such as enantioselective catalysis or separation, chiral sensing, drug delivery, and the creation of circularly polarized light (CPL).[1-8] These applications arise from the different interactions of chiral plasmonic NPs with left- and right-handed CPL. Therefore, much effort has been put into the development of NPs with complex structures and morphologies.[9-17] The properties of such samples are usually characterized by measuring their circular dichroism (CD), which quantifies the interaction of an ensemble of particles with CPL.[18] Under certain conditions, such measurements can also be performed for single nanoparticles, which may help accounting for heterogeneity in morphology and optical response.[19-21] Multiple factors can be at the origin of the recorded CD signal, such as a chiral morphology, chiral features in the crystalline structure, or the presence of chiral molecules at or near the surface of the chiral NPs.[1,7,22-25] To obtain NPs with tailored chiroptical properties, it is thus important to understand the connection between the CD signal and the NP morphology.

Electron microscopy is a suitable technique to investigate the structure of single chiral NPs. However, neither scanning electron microscopy (SEM) nor (scanning) transmission electron microscopy ((S)TEM) can faithfully reveal the complete 3D structure of the NPs.[14,26,27] A promising alternative is electron tomography (ET), a technique that allows the reconstruction of the 3D structure of nanomaterials from a tilt series of 2D STEM projection images. ET has been applied to visualize the 3D structure of chiral Au NPs in great detail.[10,28,29] For relatively simple structures, one can manually measure characteristics such as the helical pitch. Unfortunately, more complex structures, such as the chiral Au nanorods (NR) introduced by González-Rubio et al., are more difficult to quantify on the basis of a purely visual inspection.[10] Using the 3D Fourier transform (FT) of high-quality 3D reconstructions, González-Rubio et al. could identify chiral features on their NRs. However, this approach could not provide a quantitative measure of the NP chirality, or even clearly resolve their handedness.[10]

The Hausdorff chirality measure has already been used to quantify chirality.[30] This method is based on minimizing the "Hausdorff distance" between the molecule and its mirror image, where a molecule is represented by a set of points in 3D space and each point represents a single atom. The chirality measure is defined as the ratio between the minimum Hausdorff distance and the diameter of the set of points. Since the number of points required to accurately represent a NP quickly rises for complex shapes, optimization becomes exceedingly time demanding. Moreover, the Hausdorff chirality measure can quantify the degree of chirality of a given system, but is unable to distinguish between left- and right-handed chirality. Other methods have also been proposed to identify mirror symmetry, but they are all specifically designed for molecular structures and/or face the same problem of computational requirements and inability to detect handedness.[22,31-34] Recently, the chirality of elongated nanocrystals was analyzed by dividing the crystal into thin layers.[35] The orientation of each layer was determined using principal component analysis and the difference in orientation between subsequent layers was used as a measure of chirality. This is a viable approach, but it requires that the orientation of the layers in the object can be determined, which is not always possible. Therefore, we are still missing a reliable method that can objectively quantify morphological chirality in single NPs, based on ET. Such a method would be extremely useful to gain the necessary insights on structural enhancements, that can eventually be used to improve the chiroptical activity of NPs.

We present herein a method to quantitatively investigate the helicity of single NPs, based on ET reconstructions. The methodology is used to identify helical features in the NP morphology and to extract parameters such as the inclination angle of such features. This approach will thus enable a quantitative investigation of the connection between NP shape and features in the corresponding CD spectrum, eventually leading to the optimization of chiral NPs.

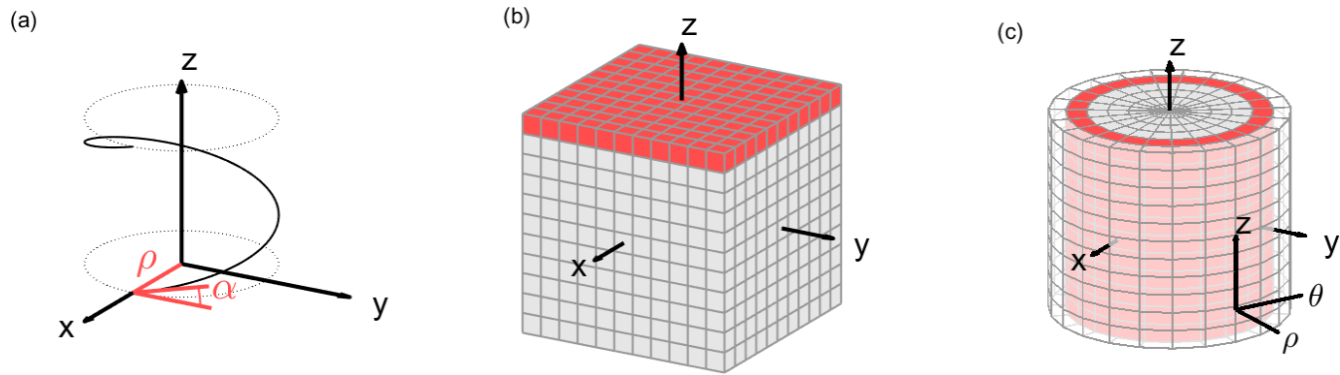

**Figure 1.** Illustration of the base geometries used in our method: (a) a helix around the z-axis with radius ρ and inclination angle α; (b) a discrete voxel grid in Cartesian coordinates, where each cube is a voxel; and (c) a discrete voxel grid in cylindrical coordinates, where the voxels are no longer cubes, but trapezoidal prisms. An orthoslice at constant z and a cylindrical section at constant ρ are highlighted in red in (b) and (c), respectively.

## Methodology

An object is chiral when its mirror image cannot be translated or rotated to completely overlap with the original shape.[36] Helicity, on the other hand, is more vaguely defined as the extent to which a structure resembles or contains helices or helix-like features. Helical shapes are always chiral but additionally present features that are easier to interpret, such as a well-defined handedness and helical pitch or inclination angle. Since most synthesis efforts towards chiral nanoparticles make use of helical growth, we designed a method to quantify helicity rather than chirality, even though chirality is the more general characteristic. Our method is specifically designed for the analysis of chiral NRs, which present attractive properties.[10,14,25,29] In a pioneering example, González-Rubio et al. were able to synthesize Au NRs with tunable chiroptical properties.[10] Beyond the initial qualitative interpretation of ET reconstructions for such particles, our method allows extracting quantitative information on chirality. Although our methodology was designed for NRs, only few assumptions on the input data were made. The method is thus relevant to most types of helical (nano)structures, as exemplified below.

The basis of our method is strongly connected to the properties of a helix, corresponding to: (i) a central axis, (ii) the distance ρ between the helix and its central axis, and (iii) the angle of inclination α (Figure 1a). A helical shape can be described as a superposition of (parts of) individual helices, such that quantifying the helicity of an object can be simplified into detecting a sum of helices. We furthermore assume that the central axis is the same for all helices within a given shape and that it can be identified manually. The central axis for NRs will be parallel to its longitudinal symmetry axis and will pass through its center of mass.

As a first step, the ET reconstruction of the object of interest is oriented such that the central axis becomes the z-axis of the defined coordinate system. Our method then searches for the presence of helices, for each combination of $\rho$ and $\alpha$. In a cylindrical coordinate system, helices around the z-axis correspond to straight lines. Since an ET reconstruction is represented on a discrete voxel grid in Cartesian coordinates, as illustrated in Figure 1b, it should therefore be converted into a discrete voxel grid in cylindrical coordinates by linear interpolation (Figure 1c). This allows us to separately investigate concentric cylindrical 2D sections at constant $\rho$, such as the red section in Figure 1c.

To illustrate the different steps of our method, we created a 3D model of a rod with a right-handed helical shell, shown in Figure 2a. The model consists of an achiral cylindrical core with radius $\rho_{core} = 64$ voxels and height $h = 256$ voxels. Wrapped around this core is a chiral shell comprising two helices (blue and red in Figure 2a), each wrapping the core 2.5 times and extending to a radius $\rho_{shell} = 96$ voxels.

Two different cylindrical sections through the model are shown in Figures 2b,c. The presence of helices in our model translates into diagonal lines in the cylindrical sections. For each cylindrical section (at different radii $\rho$) we can then compute the so-called directionality, to identify the preferred orientation of the diagonal lines. The directionality can be extracted from the gradient of the cylindrical section, which gives the direction of greatest change in intensity and the rate of change in intensity in each pixel of the cylindrical section.[37] The direction of the gradient is therefore orthogonal to the orientation of diagonal features in the cylindrical sections. The relative presence, or strength, of diagonal features in one cylindrical section can be quantified from the magnitude of the gradient. As such, one can create a histogram with inclination angles on the horizontal axis and the relative presence of features with each inclination angle on the vertical axis. Such a histogram can be filled by adding the magnitude of the gradient in each pixel to the histogram bin corresponding to the inclination angle in that pixel, as shown in Figure 2d and 2e for the two selected cylindrical sections. This process is visualized in more detail in Figure S1. The process is repeated for every cylindrical section and the results are recombined into a 2D histogram to extract the directionality $D(\rho, \alpha)$ of the 3D shape, expressed in units $[D] = [\rho]^{-1}[\alpha]^{-1}$ (Figure 2f). The directionality is normalized such that the sum over the complete 2D histogram equals one (see section 1 in the SI for more details).

The result for our model shows two strong peaks, highlighted by a green and an orange rectangle in Figure 2f. A vertical peak is present at $\alpha = 0°$ (green), which stems from the top and bottom edges of the particle. Another peak, ranging from $\alpha = 14°$ at $\rho = \rho_{core}$ to $\alpha = 10°$ at $\rho = \rho_{shell}$ (orange), corresponds to the helical shell in the model. The inclination angle for the directionality is defined in the range $[-90°, 90°]$. A helix is thus left-handed for $\alpha \in ]-90°, 0°[$, right-handed if $\alpha \in ]0°, 90°[$, and not helical when $\alpha$ equals 0° or 90°, which corresponds to horizontal or vertical features respectively. The peak marked in green is not helical, whereas the peak marked in orange is only present at positive inclination angles. We therefore conclude that the shell is right-handed helical. It should be noted that the orange peak is slightly curved, indicating different inclination angles $\alpha$ at different radii $\rho$. Indeed, in the simulated helical rod, the helices wrap around the core with a period of 102 voxels in the z-direction. Consequently, different inclination angles are required at different radii, as visible in Figure 2a.

For more complex shapes such as those investigated by González-Rubio et al., it becomes more challenging to interpret directionality plots, which are expected to contain different contributions at both positive and negative inclination angles.[10] To enable an objective interpretation, we define a helicity function $H(\rho, \alpha)$, as the difference between right- and left-handed bins in the directionality $D(\rho, \alpha)$:

$$H(\rho, \alpha) = D(\rho, \alpha) - D(\rho, -\alpha)$$

for $\alpha \in [0°, 90°]$. Consequently, the helicity function will be positive if there are more right-handed features with parameters $\rho$ and $\alpha$, or negative if there are more left-handed features. If the number of left- and right-handed features are equal, the helicity function will be zero. The helicity function for our model is shown in Figure 2g. A strong positive (right-handed) signal is present at the same inclination angles as earlier observed for the peak marked in orange in Figure 2f. Since $\alpha = 0°$ corresponds to non-helical features, the peak marked in green is not present in the helicity function because $D(\rho, 0°) = D(\rho, -0°)$.

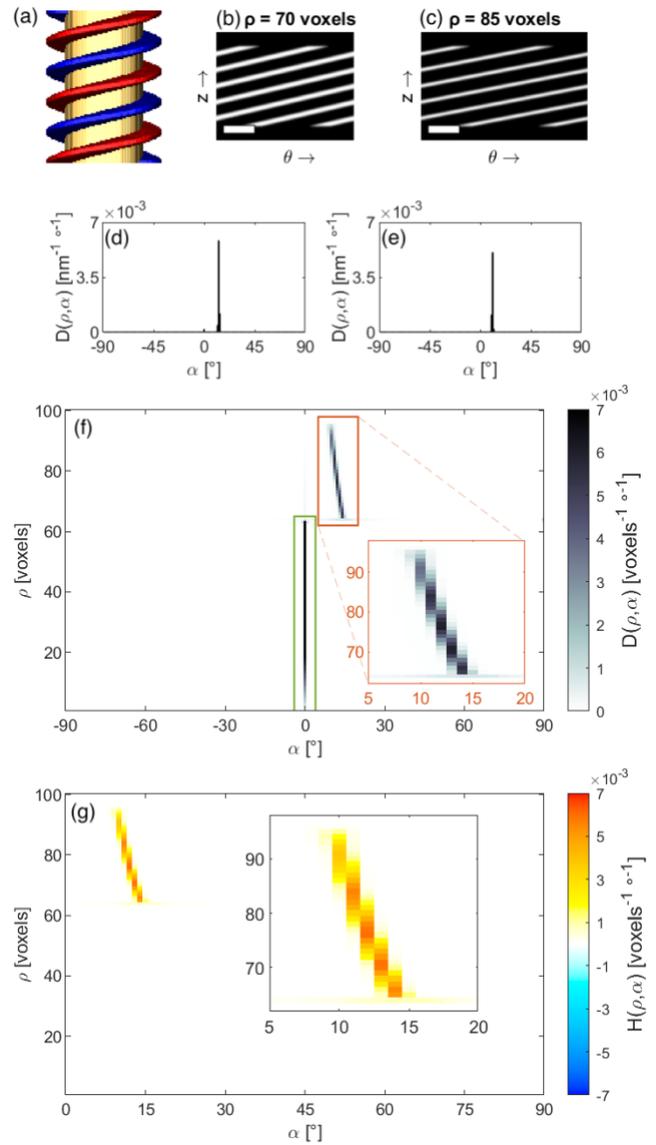

**Figure 2.** A simulated helix (a) with two of its cylindrical sections at $\rho = 70$ voxels (b) and $\rho = 85$ voxels (c), the scalebars are 100 voxels wide. The model consists of an achiral core (yellow) enveloped by two helices (blue and red). The directionality for the two cylindrical sections is shown in (d) and (e) respectively, and the directionality of all cylindrical sections is combined in (f) to obtain the directionality of the 3D simulated helix in cylindrical coordinates. Two main features in (f) are marked by green and orange rectangles, and the orange region is enlarged in the inset. (g) Helicity function $H(\rho, \alpha)$ and zoomed view (inset) of the peak corresponding to the helical shell of the model. Note that the line is not vertical due to the different angles involved in the model helix.

A single parameter that indicates the total helicity of a given NP can be useful to compare the degree of helicity for different NPs. Such a measure can be obtained by calculating the integral of the helicity function, which reduces to a sum if the helicity function is discrete:

$$H_{total} = \sum_{\rho} \sum_{\alpha} H(\rho, \alpha) \Delta \rho \Delta \alpha$$

With $H_{total} \in [-1, 1]$ due to the normalization of the directionality (see SI for more details). Positive values indicate right-handed helicity, negative values indicate left-handed helicity, and values close to zero indicate no overall helicity (even though certain parts of a NP may show helicity, this can be compensated by other areas within the same NP). More details on the calculation of $H(\rho, \alpha)$ and $H_{total}$ are presented in section 2 of the SI, where the computational efficiency of the method is also discussed (Figure S2).

**Experimental results**

**Au NRs.** To illustrate the use of our method, we investigated 3D reconstructions of Au NRs synthesized by micelle-templated seeded growth on achiral Au NRs.[10] A chiral shell was grown on the seeds using cetyl-trimethylammonium chloride (CTAC) as a surfactant, combined with either (R)-BINAMINE for right-handed chiral NRs or (S)-BINAMINE for left-handed chiral NRs. BINAMINE (1,1′-binaphthyl-2,2′-diamine) is a derivative of BINOL (1,1′-bi(2-naphthol)) and acts as a chiral co-surfactant to create helical micelles.[38] Using this approach, chiral Au NRs of different dimensions can be prepared, which may slowly reshape over time because of a limited thermodynamic stability. Therefore, the thiolated amino acid cysteine was used as a stabilizing agent to prevent reshaping. The reader is referred to the paper by González-Rubio et al. for complete details on synthesis and ET procedures.[10] Here, we selected five different NRs, as listed in Table 1. Two of the selected particles (R1 and R2) were obtained from the same sample, prepared with (R)-BINAMINE/CTAC; two other NRs (S1 and S2) were from a sample prepared with (S)-BINAMINE/CTAC. The synthesis conditions for these particles were the same, except for the specific BINAMINE enantiomer. A fifth particle (R3) was prepared with (R)-BINAMINE/CTAC, but under conditions resulting in a thinner chiral shell. More details on the synthesis are provided in section 3 of the SI.

3D renderings of the corresponding tomography reconstructions are shown in Figure 3 (see also orthoslices in Figure S3). Due to their thickness and the limited depth of field of the electron probe, only the front half of the particles could be reconstructed at a sufficient resolution for further quantitative analysis.[39] In Figure 3, the NRs are oriented to display their front halves and complete 3D rendering movies are provided as part of the SI. Interestingly, our method provides the opportunity to calculate the helicity function for selected parts of the reconstruction. We therefore selected those parts of the reconstructions that showed sufficient resolution for further analysis. This approach will not introduce bias since the particles are randomly oriented on the TEM grid. Since the tips of the chiral Au NRs are dominated by randomly oriented wrinkles, they were manually removed before calculating the helicity functions $H(\rho, \alpha)$ and the total helicity $H_{total}$, presented in Figure 3. The results including the tips are shown in Figure S4 and it is clear that the presence of random wrinkles indeed affect the results.

The $H_{total}$ values, calculated for the particles synthesized using (R)-BINAMINE/CTAC (R1, R2 and R3) are all positive, as expected for a right-handed helical structure, whereas those obtained using the (S)-enantiomer (S1 and S2) are left-handed helical (negative $H_{total}$). This result confirms the claim by González-Rubio et al. that the handedness of chiral NRs can be controlled through rational selection of the corresponding enantiomer during synthesis.[10] Moreover, the absolute value of $H_{total}$ is similar for R1, R2 and S2, as expected because they resulted from similar synthesis conditions. On the contrary, a much lower $H_{total}$ value was obtained for S1, which is likely due to a significant variability in the chirality of different NRs from the same sample batch, as also shown experimentally.[20,21]

In this particular system, the helicity of the particles originates from helical wrinkles in the shell. As discussed by González-Rubio et al., the wrinkles have inclination angles ranging from $\alpha = 0°$ to $\alpha = 45°$.[10] The plots of the helicity function $H(\rho, \alpha)$ confirm these conclusions and furthermore reveal that the larger particles (R1, R2, S1 and S2) contain mostly inclination angles between $\alpha = 0°$ and $\alpha = 25°$, with most of the helicity being concentrated around $\alpha = 10°$. On the contrary, the wrinkles for the smaller particle (R3) feature higher inclination angles, between $\alpha = 20°$ and $\alpha = 50°$. Although additional data are required to confirm this trend, the results indicate that the average angle of inclination strongly depends on the thickness of the chiral shell.

**Table 1.** Experimental details of five selected NRs used to test the method. All NRs comprise a central achiral Au NR, on which a chiral Au shell was grown using surfactant micelle templating, using either (R)-BINAMINE/CTAC or (S)-BINAMINE/CTAC mixtures.

| Particle | Micelle | Seed diameter [nm] | Seed height [nm] | Diameter [nm] | Height [nm] | Expected handedness |
|---|---|---|---|---|---|---|
| R1 | (R)-BINAMINE | 46 | 139 | 127 | 216 | Right-handed |
| R2 | (R)-BINAMINE | 48 | 143 | 131 | 223 | Right-handed |
| S1 | (S)-BINAMINE | 46 | 131 | 135 | 212 | Left-handed |
| S2 | (S)-BINAMINE | 42 | 132 | 124 | 208 | Left-handed |
| R3 | (R)-BINAMINE | 43 | 131 | 81 | 177 | Right-handed |

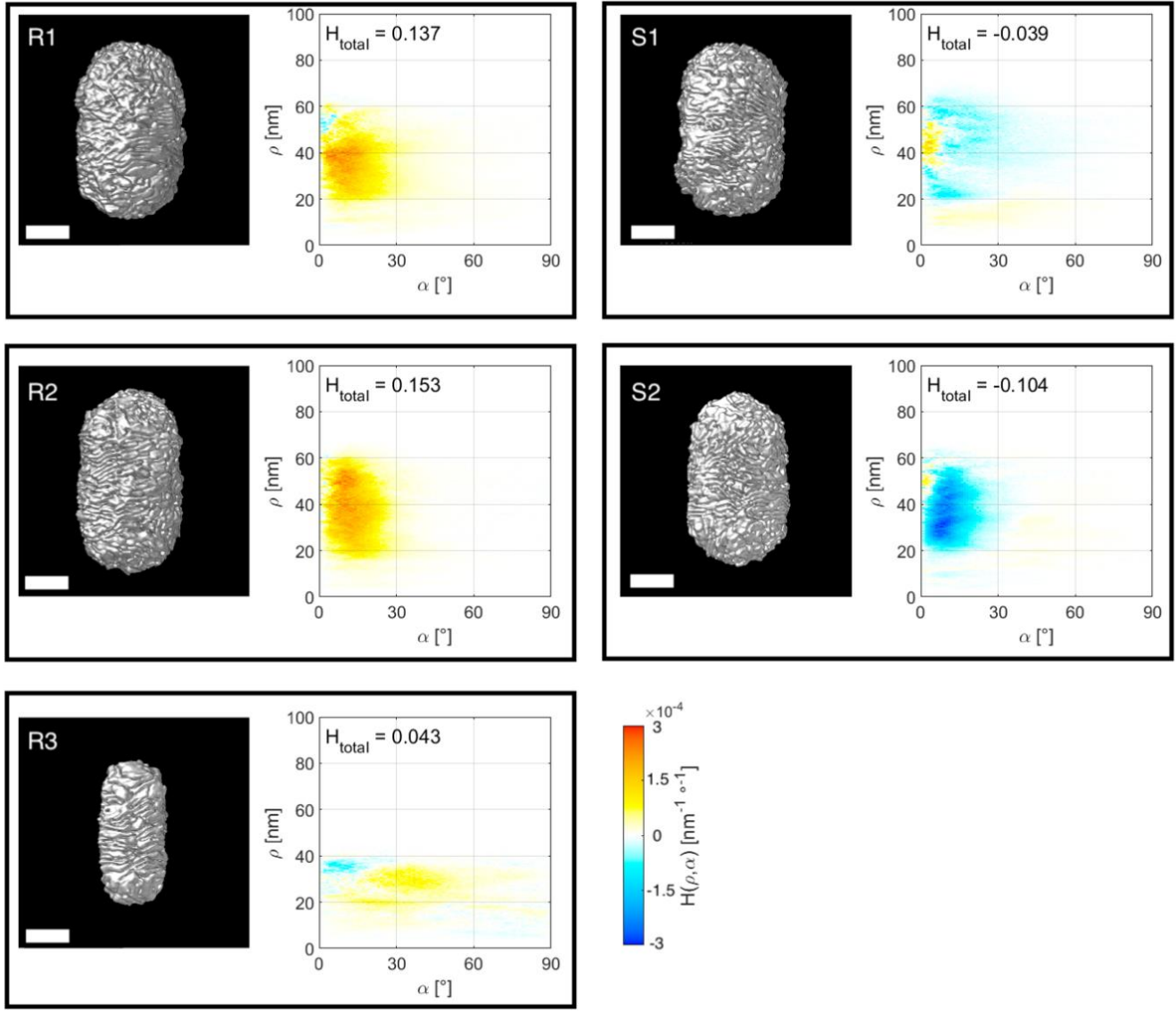

**Figure 3.** Isosurface visualization of the 3D ET reconstructions of five chiral Au NRs, described in Table 1 (left image in each panel), along with a plot of the corresponding helicity function $H(\rho, \alpha)$ (right image in each panel); The total helicity $H_{total}$ is also indicated for each Au NR. The scale bars are 50 nm.

From our analysis, it also appears that thinner chiral shells display a lower fraction with chiral signature. This observation is also reflected by $H_{total}$, which is significantly lower for R3, as compared to R1 and R2. These preliminary results agree with the average optical anisotropy factors measured experimentally for right-handed chiral NRs.[10] A maximum of $g_{CD,max} = 0.12$ at a wavelength of 700 nm was recorded for the sample containing particle R3, whereas for the sample containing R1 and R2 a maximum anisotropy factor of $g_{CD,max} = 0.18$ was obtained at a wavelength of 1100 nm. As a comparison with well-defined chiral NRs, we additionally applied our methodology to two different particles discussed in ref. 10. The first has a qualitatively different helicity function $H(\rho, \alpha)$, but is still right-handed helical, while the second particle has poorly defined chirality (Figure S5 and section 4 in the SI), which correctly results in a much lower $H_{total}$ value than for the nanorods discussed in Figure 3.

**Visualization of helical features.** A purely visual inspection of the NRs indicates that they contain both left- and right-handed features, regardless of their overall handedness. It is consequently useful to identify the helicity of subregions in the particle. We therefore created 3D helicity maps to identify which features contribute most to the overall helicity. To build these maps, we calculated the helicity measure for small windows around each voxel in the particle in cylindrical coordinates and then assigned the calculated total helicity $H_{total}$ to each voxel (Figure S6). The main drawback of this approach is that the window size must be manually selected. After comparing different window sizes, as discussed in section 5 of the SI and Figure S7, we concluded that a window size of $32 \times 32$ voxels is the optimal choice for this particular case.

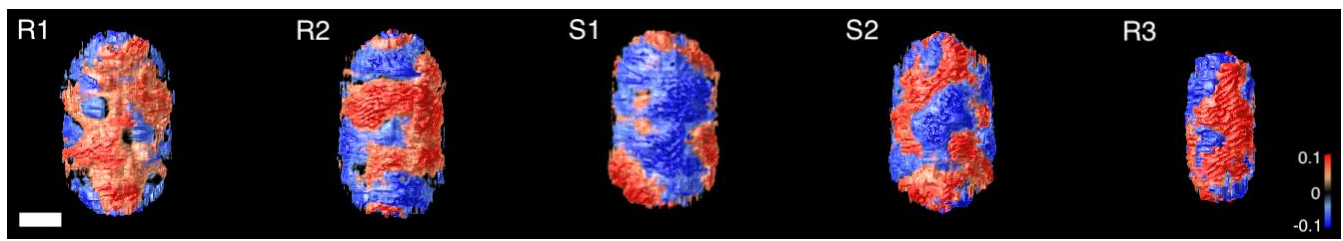

**Figure 4.** 3D color-coded volume renderings of the helicity maps for the particles described in Table 1. Red indicates right-handed helical features and blue indicates left-handedness. The scale bar (valid for all images) represents 50 nm.

Figure 4 shows the helicity maps for the NRs listed in Table 1. Orthoslices through these maps (Figure S8) and animated helicity maps are provided in the SI. As expected, all chiral NRs contained both right- and left-handed features. However, arguably more right-handed features than left-handed features can be observed in globally right-handed rods and vice versa. These results are of interest to further understand the origin of the chiral features and to eventually optimize the synthesis toward NRs with enhanced optical chirality.

**Complex helical nanostructures.** Although our methodology has been specifically designed for the analysis of helical NRs, the limited number of assumptions allow us to apply the approach to other helical systems, as long as a well-defined helical axis can be identified. E.g., we analyzed the helicity of self-assembled Au NRs around amyloid fibrils (Figure 5).[40] Such fibrils are formed by spontaneous aggregation of amyloid proteins and are known to be related to various neurodegenerative diseases. Since these fibers display a (double-)helical morphology, they could be used as a template for the helical organization of Au NRs. Under such arrangement, the coupling of plasmonic effects from Au NRs results in optical activity that can be recorded as circular dichroism (CD). Kumar et al. proposed the use of CD signals from helically assembled Au NRs to selectively detect the presence of the fibrils.

Characterization of one such self-assembled Au NR double-helical fibril by cryo-ET provided a 3D reconstruction that could be analyzed by the proposed method. Figure 5 shows that the well-defined helicity of the system resulted in a value of $H_{total} = -0.28$, which is closer to a perfect left-handed helical object than the NRs investigated above. The helicity function $H(\rho, \alpha)$ (Figure 5c) indeed shows a strong left-handed helical signal at high inclination angles, between $\alpha = 60°$ and $\alpha = 85°$. From the position of the peak in the helicity function, between $\rho = 10$ nm and $\rho = 20$ nm, we conclude that the nanoparticles are organized within this range of radii from the helical axis. In contrast to the findings for the helical NRs, the helicity map (Figure 5b) indicates that all regions of the cluster contribute to the left-handed helicity, with limited traces of right-handedness that originate from a few misaligned NRs.

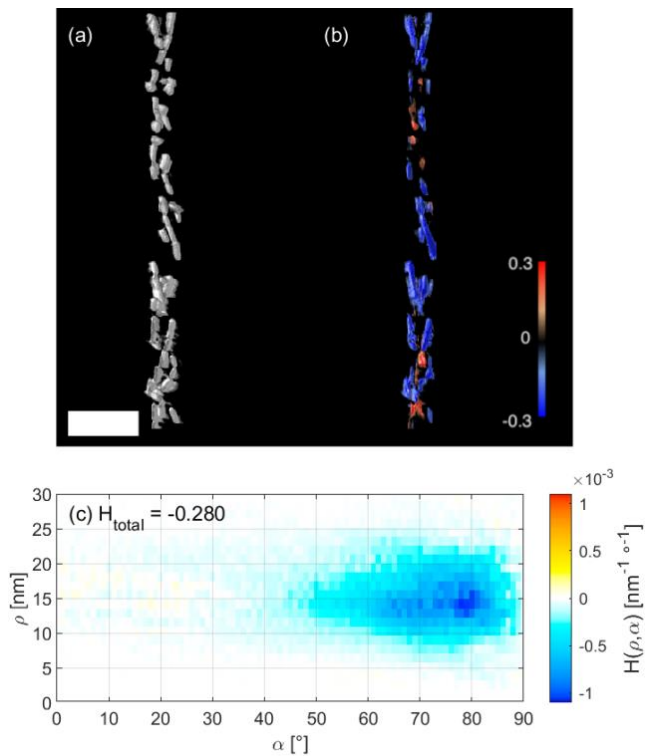

**Figure 5.** Quantitative analysis of the helicity of a helical arrangement of Au NRs around an amyloid fibril. An isosurface rendering of the NRs is shown in (a), the helicity map is shown in (b) as a semi-transparent 3D color-coded volume rendering, where red indicates right-handed helicity and blue indicates left-handed helicity. The helicity function $H(\rho, \alpha)$ and total helicity $H_{total}$ are given in (c). The scale bar is 100 nm.

**Discussion and outlook**

Our method, for which a Python package (HeliQ) was created,[41] is complementary to ensemble measurements and yields a wealth of information at the single particle level. However, a few limitations must be overcome to make the approach widely applicable to any type of system. First, the helical axis must be identified manually, which currently restricts the use of our method to NRs, nanowires, and similar elongated structures. This became apparent when analyzing the NRs in Figure 3, where the seemingly random orientations of the wrinkles at the tips of the NRs were found to introduce artefacts in the results. Namely, one could hypothesize that the wrinkles at the tips are also helical, but with a different helical axis. A future possibility might be to use a 3D gradient to (locally) detect the helical axis. Another limitation lies in the current normalization of the directionality. If only few features are present in a cylindrical section (i.e., if it has a nearly uniform intensity), then the normalization factor will be small, which artificially increases the intensity in the helicity function $H(\rho, \alpha)$. An example of this behavior is shown in the achiral NR in Figure S5. As discussed in the SI, this mainly forms a problem for non-helical axially symmetric particles. A different normalization factor could be the solution for this problem. Nonetheless, the total helicity $H_{total}$ accurately identifies the second particle in Figure S5 as achiral. Taking the current limitations into account, future modifications of our methodology will likely make it generally applicable to any shape.

## Conclusion

We introduced a method for the quantitative analysis of morphological helicity, which has been exemplarily applied to ET reconstructions of chiral Au nanorods. The approach is based on the geometrical properties of a helix and results in a two-dimensional helicity function $\mathbf{H(\rho, \alpha)}$, which can be interpreted as the decomposition of a given shape into a combination of helices. The helicity function provides spatially resolved information about the presence of helical features, as well as their inclination angles. A numerical parameter, the total helicity $\mathbf{H_{total}}$ obtained as the integral over the full helicity function, gives an overall indication of the helical degree of a given particle. Analyses of ET reconstructions of chiral Au NRs agree with previous observations, while additionally providing more quantitative parameters, such as the angle of inclination of helical features. This information may eventually help understand the effect of various synthesis parameters. The more general applicability of the approach was demonstrated through the analysis of a helical nanostructure composed of Au NRs clustered around a fibrillar protein template; a high degree of helicity throughout the entire super-helical nanostructure was confirmed, in agreement with previous experimental measurements.

## ASSOCIATED CONTENT

### Supporting Information

The Supporting Information is available free of charge on the ACS Publications website.

Additional images, synthesis and computational details (PDF)

Supporting movies visualizing the connection between particle morphology and helicity maps (MP4)

## AUTHOR INFORMATION

### Corresponding Authors
*E-mail: Sara.Bals@uantwerpen.be.
*E-mail: W.Albrecht@amolf.nl.

### Author Contributions
The manuscript was written through contributions of all authors. All authors have given approval to the final version of the manuscript.

## ACKNOWLEDGMENTS

S. B. and A. P.-T. gratefully acknowledge funding by the European Research Council (ERC Consolidator Grant #815128-REALNANO). L. M. L.-M. acknowledges funding from the Spanish Ministerio de Ciencia e Innovación through grant # PID2020-117779RB-I00 and the Maria de Maeztu Units of Excellence Program from the Spanish State Research Agency (Grant No. MDM-2017-0720). G. G.-R. thanks the Spanish Spanish Ministerio de Ciencia e Innovación for FPI (BES-2014-068972) fellowship.